\documentclass[twocolumn,english,aps,prb]{revtex4}

\usepackage{mathrsfs}
\usepackage{amstext}
\usepackage{graphicx}
\usepackage{babel}
\usepackage{epstopdf}

\makeatletter

\newcommand{\Rmnum}[1]{\expandafter\@slowromancap\romannumeral #1@}
\makeatother

\begin{document}

\title{The single-electron transport in a three-ion magnetic molecule
  modulated by a transverse field}

\author{Javier I. Romero}

\email{jromerobohorquez@knights.ucf.edu}

\affiliation{Department of Physics, University of Central Florida,
  Orlando, Florida 32816, USA}

\author{Eduardo R. Mucciolo}

\email{mucciolo@physics.ucf.edu}

\affiliation{Department of Physics, University of Central Florida,
  Orlando, Florida 32816, USA}

\date{\today}

\begin{abstract}
We study single-electron transport in a three-ion molecule with strong
uniaxial anisotropy and in the presence of a transverse magnetic
field. Two magnetic ions are connected to each other through a third,
nonmagnetic ion. The magnetic ions are coupled to ideal metallic leads
and a back gate voltage is applied to the molecule, forming a
field-effect transistor. The microscopic Hamiltonian describing this
system includes inter-ion hopping, on-site repulsions, and magnetic
anisotropies.
For a range of values of the parameters of the Hamiltonian, we obtain
an energy spectrum similar to that of single-molecule magnets in the
giant-spin approximation where the two states with maximum spin
projection along the uniaxial anisotropy axis are well separated from
other states. In addition, upon applying an external in-plane magnetic
field, the energy gap between the ground and first excited states of
the molecule oscillates, going to zero at certain special values of
the field, in analogy to the diabolical points resulting from Berry
phase interference in the giant spin model. Thus, our microscopic
model provides the same phenomenological behavior expected from the
giant spin model of a single-molecule magnet but with direct access to
the internal structure of the molecule, thus making it more
appropriate for realistic electronic transport studies.
To illustrate this point, the nonlinear electronic transport in the
sequential tunneling regime is evaluated for values of the field near
these degeneracy points. We show that the existence of these points
has a clear signature in the $I-V$ characteristics of the molecule,
most notably the modulation of excitation lines in the differential
conductance.
\end{abstract}

\maketitle

\section{Introduction}

From the various types of magnetic states in matter, ferromagnetism
and its microscopic causes is one of the most intriguing topics. A
very singular class of magnetic systems are single-molecule magnets
(SMMs).\cite{christou2000,gatteschi2007,friedman2010} Molecules in
this class typically have a large net spin ground state and exhibit
unusual attributes such as quantum tunneling of the magnetization
(QTM),\cite{friedman1996,thomas1996} a relatively large decoherence
time,\cite{ardavan2007} and Berry-phase interference effects in the
presence of magnetic fields.\cite{wernsdorfer1999,quddusi2011}
Commonly, SMMs are composed of transition metal ions (open $3d$ or
4$f$ shells) bridged by ligand atoms and molecules. The
phenomenological Hamiltonian that describes the magnetic properties of
SMMs is the so-called giant spin approximation (GSA) model. For
instance, in its simplest form, the GSA Hamiltonian can be written as
\begin{equation}
\label{eq:1}
H_{\textrm{GSA}} = -D\, S_{z}^{2} + E\, \left( S_{x}^{2} -S_{y}^{2}
\right),
\end{equation}
The net magnetization has a preferential direction (an easy axis $z$
in this example) as a consequence of the dominant uniaxial anisotropy
($D \ll E$). The in-plane transverse anisotropy allows the total spin
to transit between different values of $S_z$. In this particular
example, the molecule has a predominant second-order anisotropy
signaling a rhombic symmetry.
 
A more microscopic description of SMMs comes from considering
interactions at the ion level. In this case, one uses instead a
multi-spin Hamiltonian of the type
\begin{equation}
\label{eq:2}
H_{\textrm{ms}} = \sum_{i} \left[ -d_{i} S_{iz}^{2} + e_{i}
  \left( S_{ix}^{2} - S_{iy}^{2} \right) \right] - \sum_{i\ne j} J_{ij}
\vec{S}_{i} \cdot \vec{S}_{j}.
\end{equation}
Here each magnetic site of the molecule has local uniaxial and
transverse anisotropies $d_{i}$ and $e_{i}$, respectively. In
addition, there is an effective isotropic ferromagnetic interaction
between pairs of sites parametrized by $J_{ij}>0$, which contributes
to the energy splitting of states with different total spin. (Some
molecules are better described by an anisotropic $J_{ij}$.) We note
that the on-site anisotropy terms (proportional to $d_{i}$ and
$e_{i}$) in Eq. (\ref{eq:2}) are meaningful only when the total spin
of each site is $S_{i}\geq 1$.

The accuracy of the GSA and the multi-spin models was investigated in
Ref. \onlinecite{Hill2010} by studying the connection between
molecular and single-ion uniaxial anisotropies. Anisotropies in SMMs
are very sensitive to the orientation of the ligands with respect to
each magnetic ion, thus the energy spectrum of SMMs varies greatly
with magnetic site symmetries. For instance, in the case of zero
transverse anisotropies ($e_{i} = 0$ and, consequently, $E = 0$), the
Hamiltonians in Eqs. (\ref{eq:1}) and (\ref{eq:2}) yield a two-fold
degenerate ground state involving spin states parallel ($S_{z}=S$) and
anti-parallel ($S_{z} = -S$) to the uniaxial anisotropy axis. If the
transverse anisotropy terms ($e_{i}$ and $E$) are nonzero, then the
rotational $S_{z}$ symmetry is broken and the two-fold degeneracy is
lifted, with the ground and first excited states now being
antisymmetric and symmetric combinations of the $S_{z} = \pm S$
states. In general, the parameters used in the multi-spin approach
depend on the intra-site or inter-site interactions between orbitals
of the ligands and transition metals, but due to the large Hilbert
space involved and the strong interplay between the many microscopic
parameters, it is very challenging to take into account explicitly the
overlap between orbitals and their symmetries for large magnetic
molecules. One often employs the empirical Goodenough-Kanamori rules
\cite{goodenough1955,kanamori1959} that dictate the nature of
interaction between magnetic ions. In addition, one also includes an
effective interaction intermediated by diamagnetic atoms (the Anderson
superexchange interaction).\cite{anderson1950}

In recent years, interest emerged in exploring how the magnetic
properties of SMMs may affect the molecule's electronic transport
properties. Effects due to the QTM,\cite{romeike2006a} Berry-phase
interference,\cite{gonzalez2006,leuenberger2006} and their interplay
with the Kondo resonance \cite{romeike2006b,gonzalez2008} have been
proposed. Several authors have also used first-principles calculations
to investigate the electronic configuration, the magnetic properties,
and coherent transport in
SMMs.\cite{park2005,barrazalopez2009,park2010,nossa2012} In addition,
many important contributions have been made in studying the sequential
transport for different molecular setups,\cite{Elste,Lehmann1} as well
as sequential transport dependence on spin-orbit coupling, the
Jahn-Teller effect, and ligand charge
variations.\cite{Herzog,Reckermann,Romeike2007}

In this paper, we investigate the dependence of sequential tunneling
transport on an transverse magnetic field applied to a three-ion model
that contains the essential microscopic details necessary for
reproducing a SMM behavior. The microscopic model comprises two
magnetic ions bridged by a third diamagnetic ion and takes into
account the valence, ligand fields, and orbital energies of the ions,
as well as direct and exchange Coulomb interactions present in the
molecule. Our goal is to develop a simple phenomenological
microscopic model of a SMM that goes beyond the giant spin
Hamiltonian model and is amenable to realistic electronic transport
studies.

We explore the energy spectrum of this system for several points in
the parameter space of the Hamiltonian. We find that for a certain
parameter range a magnetic bistability, similar to that observed for
SMMs, develops. This bistability is characterized by a ground state
degeneracy point and is driven by a magnetic field perpendicular to
the molecule's main anisotropy axis. The modulation of the transition
between magnetic states through a transverse magnetic field is one of
the most unusual features of SMMs and is fully reproduced by our
model. In the context of the giant spin model of an SMM, this
modulation is understood as the result of Berry-phase interference of
multiple spin tunneling paths, which in turn lead to the appearance of
diabolical points in the molecule's spectrum.\cite{vondelft+garg} In
contrast, Our model contains many degrees of freedom and there are no
easily identified topological phases. Instead, the modulation in our
model arises from changes in electronic correlations when a finite
transverse field is present. We evaluate the incoherent electronic
transport through the molecule near this bistability. At the
degeneracy points, states with opposite spins are decoupled, resulting
in a clear qualitative change in the differential conductance of the
molecule.

The paper is organized as follows: In Sec. \ref{sec:model} we describe
the model Hamiltonian for the three-ion magnetic molecule. In Sec.
\ref{sec:parameters} we present our choice of model parameters and the
energy spectrum and symmetry of the states of the molecule. In
Sec. \ref{sec:transport} we evaluate the sequential current and
differential conductance of the molecule for in-plane magnetic fields
ranging between zero and the nearest degeneracy point. In
Sec. \ref{sec:modulation} we discuss the effects of spin tunneling
modulation by a magnetic field on transport. Finally, in
Sec. \ref{sec:conclusion}, we analyze our results and draw some
conclusions.

\begin{figure}[ht]
\includegraphics[width=0.80\columnwidth]{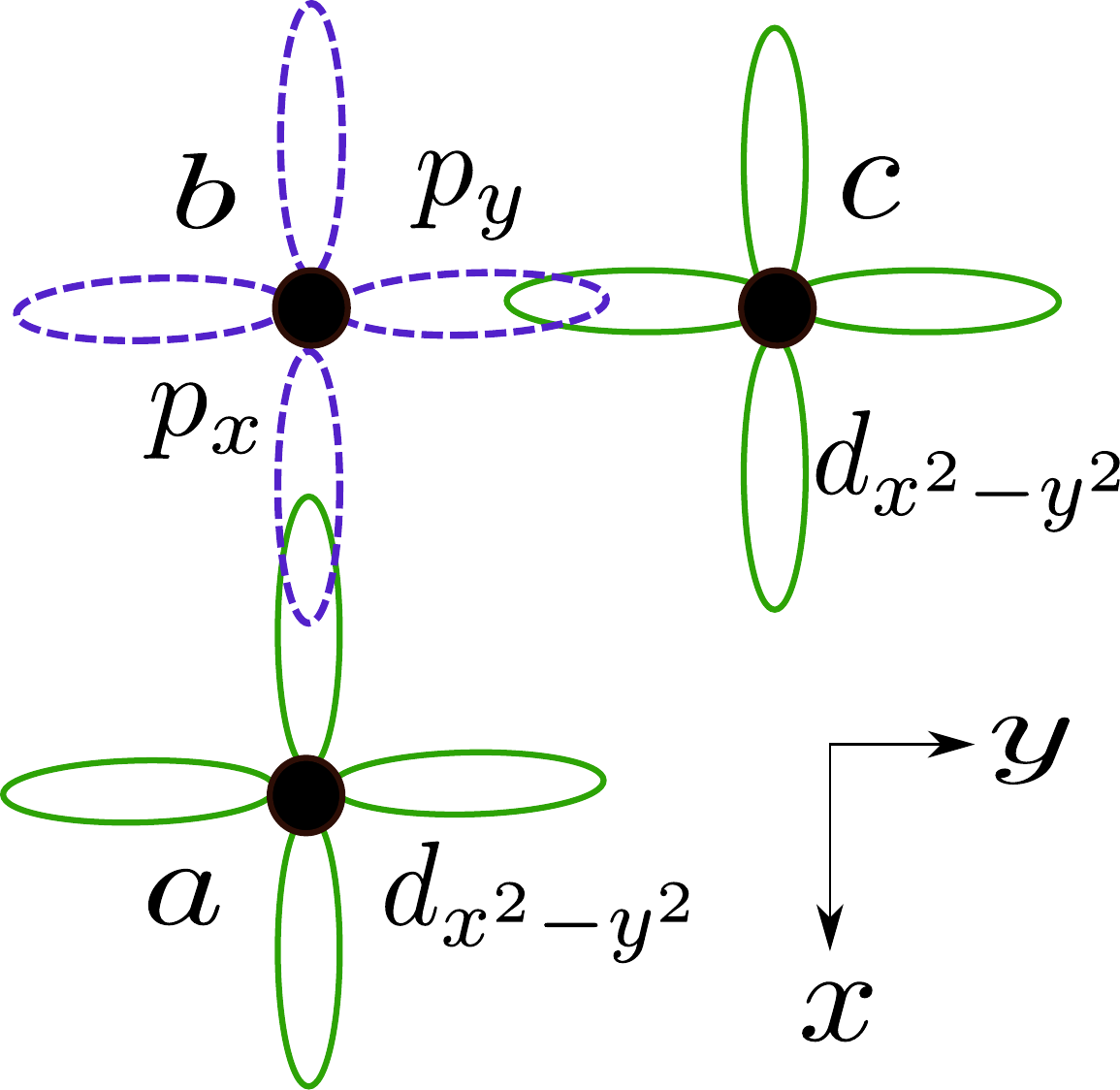}
\caption{(Color online) Scheme of a simple three-ion SMM. For
  simplicity, we do not depict the rhombic ligand environment
  surrounding ions $a$ and $c$. The magnetic ions labeled by $a$ and
  $c$ interact through the diamagnetic ligand ion $b$ with a bond
  angle of $\theta = \pi/2$, so that the $d_{x^{2}-y^{2}}$ orbitals
  (green/solid lobes) overlap with the $p_{x}$ and $p_{y}$ orbitals
  (purple/dashed lobes) of the ligand separately.}
\label{fig:1}
\end{figure}

\section{The three-ion model of a SMM}
\label{sec:model}

A minimal realistic molecular core capable of reproducing the main
features of a SMM such as large total spin ($S>1/2$), uniaxial and
in-plane anisotropies, and transverse field modulation consists of two
transition metal ions bridged by a diamagnetic ligand ion. Consider
the system shown schematically in Fig. \ref{fig:1}. The two transition
metal ions $a$ and $c$ have a $3d^{8}$ electronic configuration and a
total spin $S=1$ each. They interact magnetically through a
superexchange interaction intermediated by a diamagnetic
$\textrm{O}^{2-}$ ion (electronic configuration $2p^{6}$), represented
by $b$ in Fig. \ref{fig:1}. The five-fold degeneracy of the $3d$
orbitals in the magnetic ions is broken due to the bonding to
ligands. To simulate such an effect, we assume that a weak
orthorhombic ligand field acts on $a$ and $c$, inducing local uniaxial
and transverse anisotropies on each site. Thus, for an orthorhombic
symmetry ($D_{2h}$ point group), the ground states of ions $a$ and $c$
have two occupied unpaired single-particle orbitals ($a_{1,2}$ and
$c_{1,2}$, respectively) which are symmetric and antisymmetric
combinations of $d_{x^{2}-y^{2}}$ and $d_{3z^{2}-r^{2}}$ $d$
states. We also consider a 90$^\textrm{o}$ angle between their
bonds. As a result, the $d_{x^{2}-y^{2}}$ components of the
single-particle orbitals in both ions $a$ and $c$ overlap separately
with the $p_{x,y}$ orbitals $b_{1,2}$ of ion $b$. Under this
assumption, we use the appropriate Slater-Koster two-center integrals,
which then yields $E_{p_{x},d_{x^2-y^2}} = E_{p_{y},d_{x^2-y^2}} =
(pd\sigma)\sqrt{3}/2$. This configuration favors a ferromagnetic
superexchange interaction. In our case the $E_{p_{z},d_{3z^2-r^2}}$
and $E_{p_{x,y},d_{x^2-y^2}}$ two-center integrals are zero.

\begin{figure}[ht]
\includegraphics[width=0.65\columnwidth]{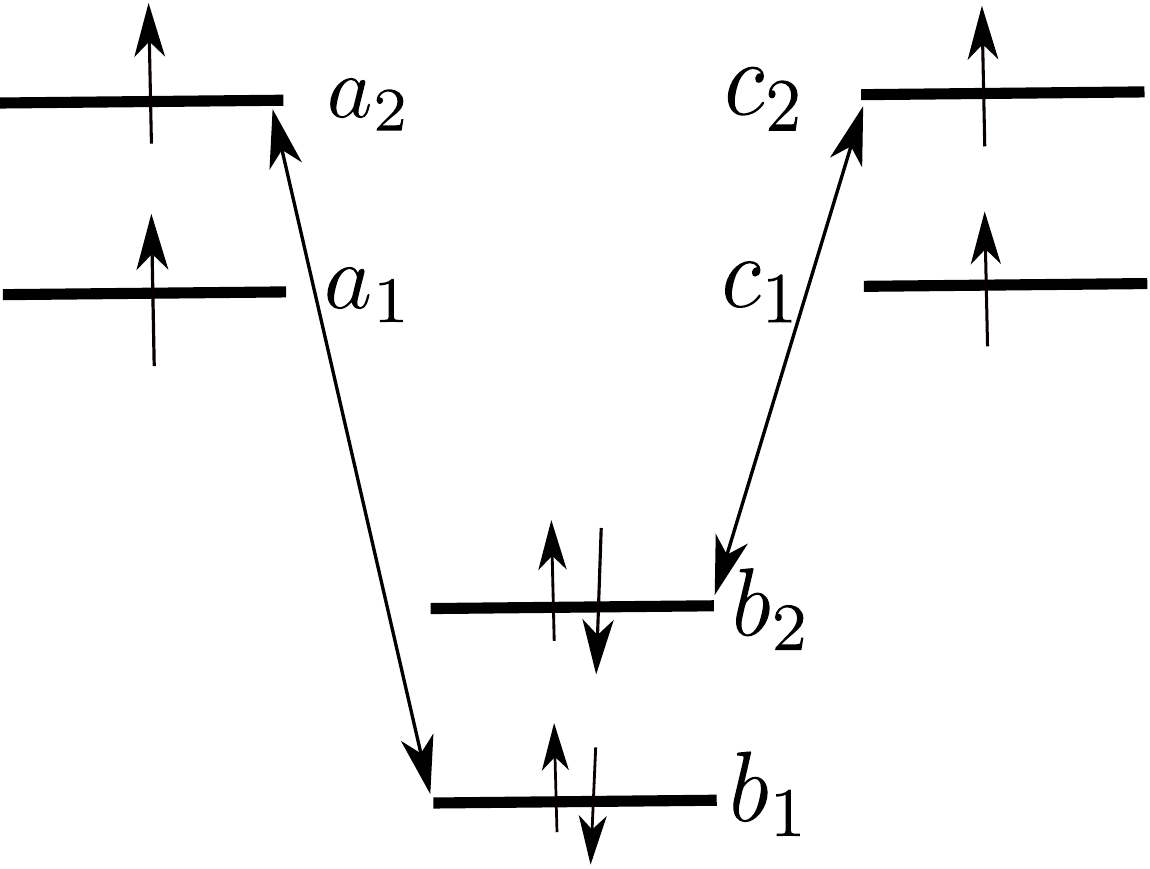}
\caption{Scheme of the electron hopping (double-arrow lines) between
  magnetic ions and the diamagnetic ion. For a $3d^{8}$ ion in a weak
  orthorhombic ligand field (i.e., with a small distortion to rhombic
  symmetry), the unpaired single-particle orbitals (say for ion $a$)
  are $\psi_{a_2} = \kappa_{1}\, d_{x^{2}-y^{2}} + \kappa_{2}\,
  d_{3z^{2}-r^{2}}$ and $\psi_{a_1} = \kappa_{1}\, d_{3z^{2}-r^{2}} -
  \kappa_{2}\, d_{x^{2}-y^{2}}$, where $\kappa_1^2 + \kappa_2^2 =
  1$. Since the rhombic contribution to the ligand field is small, we
  regard the mixing amplitude $\kappa_{2}$ between the initially
  degenerate unpaired orbitals as a small parameter: $|\kappa_2| \ll
  |\kappa_1|$.}
\label{fig:2}
\end{figure}

The scheme of these selective orbital overlaps is shown in
Fig. \ref{fig:2}. The features of our model are captured by the
effective Hamiltonian
\begin{equation}
\label{eq:Hmolecule}
H_{\textrm{mol}} = H_{a} + H_{b} + H_{c} + H_{ab} + H_{bc},
\end{equation}
where $H_{\alpha}$, with $\alpha=a_,c$ denotes the Hamiltonian for a
magnetic ion,
\begin{eqnarray}
\label{eq:Halpha}
H_{\alpha} & = & \sum_{i=1,2} \varepsilon_{{\rm M},i}\,
n_{\alpha_i,\sigma} + U_{\rm M} \sum_{i=1,2}
\sum_{\sigma=\uparrow,\downarrow} n_{\alpha_i,\uparrow}\,
n_{\alpha_i,\downarrow} \nonumber \\ & & +\ U_{\rm M}^\prime
\sum_{\sigma,\sigma^\prime=\uparrow,\downarrow} n_{\alpha_1,\sigma}\,
n_{\alpha_2,\sigma^\prime} - J_{\rm M}\, \vec{S}_{\alpha_1} \cdot
\vec{S}_{\alpha_2} \nonumber \\ & & -\ d\, S^2_{z,\alpha} + e \left(
S^2_{x,\alpha} - S^2_{y,\alpha} \right),
\end{eqnarray}
and $H_b$ describes the diamagnetic ion,
\begin{eqnarray}
\label{eq:Hb}
H_{b} & = & \varepsilon_{\rm O} \sum_{i=1,2} n_{b_i,\sigma} + U_{\rm
  O} \sum_{i=1,2} \sum_{\sigma=\uparrow,\downarrow} n_{b_i,\uparrow}\,
n_{b_i,\downarrow} \nonumber \\ & & +\ U_{\rm O}^\prime
\sum_{\sigma,\sigma^\prime=\uparrow,\downarrow} n_{b_1,\sigma}\,
n_{b_2,\sigma^\prime} - J_{\rm O}\, \vec{S}_{b_1} \cdot \vec{S}_{b_2}.
\end{eqnarray}
For simplicity we consider that the $a$ and $c$ ions have identical
rhombic symmetries and therefore they have the same interaction
strengths, orbital energies, and anisotropy parameters. We allow for a
crystal field splitting of the magnetic ion orbitals, but assume that
the orbitals in the diamagnetic ion are degenerate. In
Eq. (\ref{eq:Hmolecule}), $H_{ab}$ and $H_{bc}$ describe electron
hopping between the magnetic ions and the diamagnetic ion,
\begin{equation}
\label{eq:Hab}
H_{ab}= t \sum_{\sigma=\uparrow,\downarrow} \left (
c_{a_2,\sigma}^\dagger\, c_{b_1,\sigma} + \textrm{H.c.} \right)
\end{equation}
and
\begin{equation}
\label{eq:Hbc}
H_{bc} = t \sum_{\sigma=\uparrow,\downarrow} \left (
c_{b_2,\sigma}^\dagger\, c_{c_2,\sigma} + \textrm{H.c.} \right),
\end{equation}
with $\alpha=a,c$. In Eqs. (\ref{eq:Halpha}), (\ref{eq:Hab}), and
(\ref{eq:Hbc}), $n_{\alpha_i,\sigma} = c_{\alpha_i,\sigma}^\dagger
c_{\alpha_i\sigma}$, where $c_{\alpha_i,\sigma}^\dagger$
($c_{\alpha_i\sigma}$) creates (annihilates) an electron with $z$ spin
projection $\sigma$ in the orbital $i$ of ion $\alpha=a,b,c$ and
satisfy the standard fermionic anticommutation relations. The total
spin operator associated to an orbital $i$ in ion $\alpha$ is
$\vec{S}_{\alpha_i}$ while $\vec{S}_{\alpha} = \vec{S}_{\alpha_1} +
\vec{S}_{\alpha_2}$.

The first term on the r.h.s of Eqs. (\ref{eq:Halpha}) and
(\ref{eq:Halpha}) accounts for the single-particle orbital energies,
while the second and third terms represent the on-site and intra-site
Coulomb repulsion within ion $\alpha$. The fourth term enforces Hund's
first rule, maximizing the total spin on each ion ($J_\alpha >
0$). The last two terms on the r.h.s. of Eq. (\ref{eq:Halpha}) are the
uniaxial and transverse on-site anisotropies produced by the rhombic
ligand field environment and by the spin-orbit coupling within each
magnetic ion.

\section{Adjusting the model parameters}
\label{sec:parameters}

The model Hamiltonian in Eq. (\ref{eq:Hmolecule}) has a large number
of parameters that have to be properly adjusted in order to produce
the phenomenology expected of a SMM. Below, we discuss our choices of
parameter values. We first consider the Hamiltonian in the absence of
transverse anisotropies ($e_\alpha= 0$). Then we include non-zero
$e_\alpha $ terms, and an in-plane external magnetic field in order to
produce degeneracy points in the spectrum of the molecule.

\subsection{Total $S_{z}$ spin in the absence of transverse anisotropy}

The lowest energy states of a SMM typically have a large total spin
$S$, with the maximum spin projection $|S_z|=S$ happening at the
ground state, followed by excited states corresponding to decreasing
spin projections. A convenient way to produce such a spectrum in our
model is to start with $S_{z}$ as a good quantum number by setting
$e_\alpha = 0$. Then, for different choices of the parameters
$\varepsilon_{M,i}$, $\varepsilon_{\rm O}$, $U_{\rm M}$, $U_{\rm
  M}^\prime$, $U_{\rm O}$, $U_{\rm O}^\prime$, $J_{\rm M}$, $J_{\rm
  O}$ and $d$, we calculate the total $S_{z}$ spin of the molecule as
a function of $t$. For the states with maximum possible spin, $S=2$
for our three-ion model, we compare their composition to that expected
for a state obtained by adding two $S=1$ spin states. Even though
$U_{\rm M}^{\prime}$ and $U_{\rm O}^{\prime}$ are nonzero in a real
molecule, we have found that they have little qualitative importance in our
results. Thus, for the sake of simplicity and in order to decrease our
parameter space, we have set them equal to zero.

Realistic values for some parameters can be obtained from the review
by Imada, Fujimori, and Tokura \cite{imada1998} and the recent work by
Kim and Min \cite{kim2011} using NiO systems as a reference. Thus,
from the data in Table III of Ref. \onlinecite{imada1998} we set the
on-site Coulomb repulsion between d electrons to $U_{\rm M} = 7$ eV.
Typical values for the on-site Coulomb repulsion of p-electrons are in
the range from $4-7$ eV (see Refs. \onlinecite{Yushan1999} and
\onlinecite{Anisimov}). Here, we assume $U_{\rm O}=4$ eV. From table
III of Ref. \onlinecite{imada1998}, the Hund's rule parameter for the
magnetic ion is set to $J_{\rm M} = 0.95$ eV. For the diamagnetic ion,
we choose a slightly larger value, $J_{\rm O} = 1.5$ eV in order yield
a spectrum similar to that of a ferromagnetic SMM. The orbital
energies in the magnetic ions are obtained from Fig. 59 of
Ref. \onlinecite{imada1998}: $\varepsilon_{1}^{\rm M}-\varepsilon_{\rm
  O} = 0.66$ eV and $\varepsilon_{2}^{\rm M} - \varepsilon_{\rm O} =
0.72$ eV. For convenience, we shift the total energy such that
$\varepsilon_{\rm O}=0$. The energy splitting between the two orbitals
of the magnetic ions, $\varepsilon_{2}^{\rm M} - \varepsilon_{1}^{\rm
  O} = \Delta \varepsilon_{\rm crystal}$ is due to the crystal field
splitting produced by the surrounding ligands. In
Ref. \onlinecite{Bose1963}, rhombic crystal field parameters are found
to be of the order of meV or smaller; we choose the splitting to be
about 10 meV.

\begin{figure}[ht]
\includegraphics[width=0.9 \columnwidth]{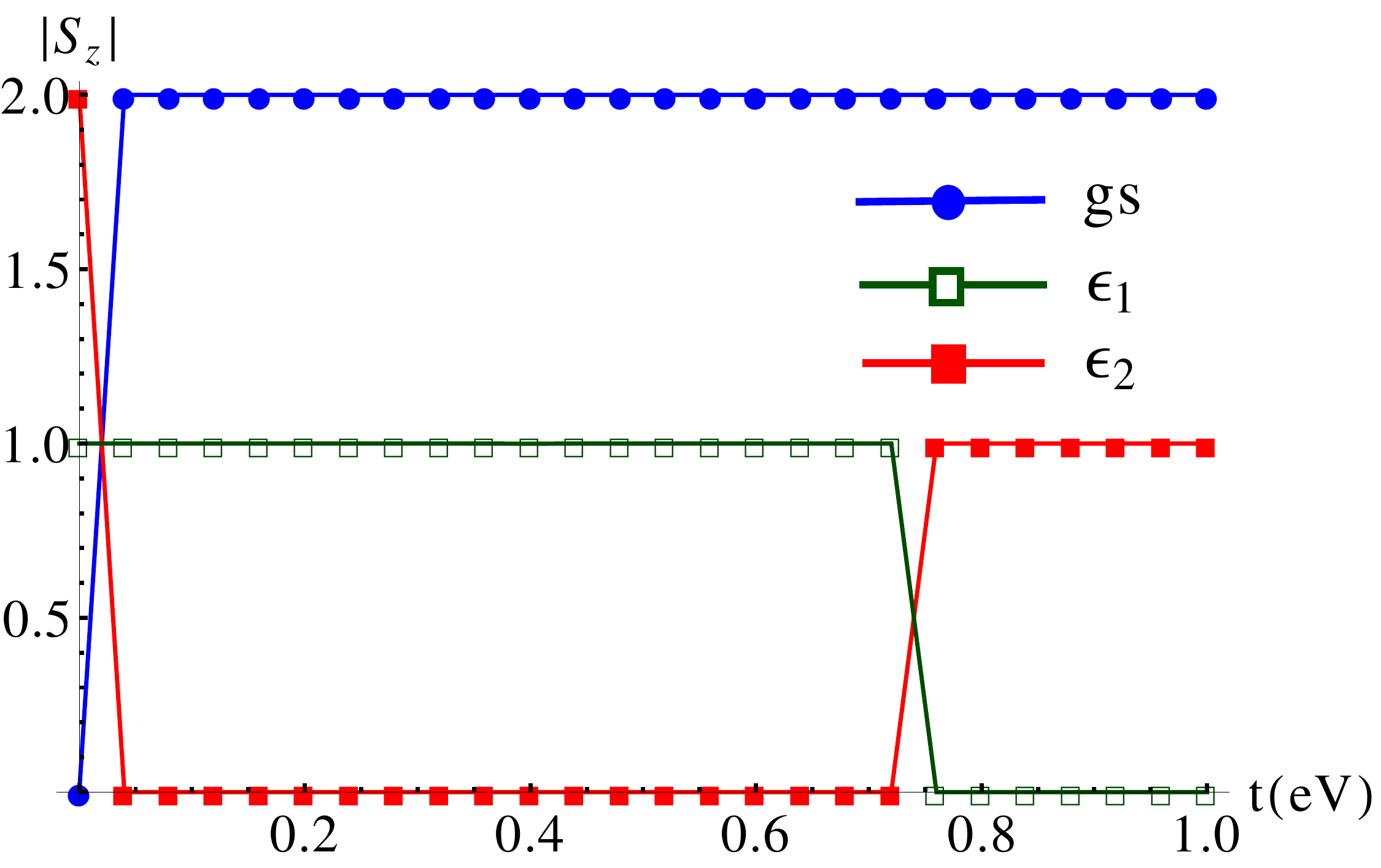}
\caption{(Color online) Total spin projection $S_{z}$ of the
  molecule. For $0 < t < 1.2$ eV, the lowest energy level is a $\vert
  \pm 2 \rangle$ state, the first excited state has $S_{z}=0$, while
  the next (degenerate) excited states have $S_{z} = \pm 1$. We find
  the typical SMM behavior in the $0.7\, \textrm{eV} < t < 1$ eV
  range, where the first excited states have $S_{z} = \pm 1$ and the
  highest energy level has $S_{z} = 0$. The crossing in the graph
  denotes the point where total $S_{z}$ spin changes for the first and
  second excited eigenstates. For $t=0.77$ eV,the energy splitting
  between the two lowest eigenstates ($S_{z}=\pm 2$ and $S_{z}=\pm 1$)
  is of the order of the uniaxial anisotropy parameters $d_{\alpha}$,
  with the value of $\Delta=0.07 $eV, while the splitting between
  $S_{z}=\pm 1$ and $S_{z}= 0$ is $\Delta=0.5$ meV.}
\label{fig:3}
\end{figure}

Based on the EPR measurements in Ref. \onlinecite{Hill2010}, the
uniaxial anisotropy parameter is set to $d = 0.6$ meV (equivalent
to 7 K). Figure \ref{fig:3} shows $\vert S_{z} \vert$ for the ground
state (two-fold degenerate), first excited state (two-fold
degenerate), and second excited state (nondegenerate) versus the
hopping parameter $t$. It is clear that for the parameter we chose,
the ground state has the highest possible total spin, $S=2$. For $t >
0.75$ eV, states with total spin projection $|S_z|=1$ become the first
excited states, while the state with total spin projection zero moves
up to the second excited state position. This is the typical case for
a SMM within the description provided by the GSA Hamiltonian of
Eq. (\ref{eq:1}).

\subsection{Anisotropy and degeneracy points}

We now consider the inclusion of local transverse anisotropies that
make the eigenstates of the Hamiltonian to be symmetric and
antisymmetric combinations of $\vert S_{z} \rangle$ states. In
addition, we consider the effect of an applied transverse (in-plane)
magnetic field on the molecule's energy spectrum. The transverse field
Hamiltonian is given by
\begin{equation}
\label{eq:26}
H_{\textrm{field}} = b \left( \sum_{\alpha=a,c} g_{x,\alpha}
S_{x,\alpha} \cos \phi + g_{y,\alpha}\, S_{y,\alpha} \sin \phi
\right),
\end{equation}
where $b = \mu_{B} \vert \vec{B} \vert$ and $\vec{B}=(B \cos\phi, B
\sin\phi,0)$ is the external magnetic field. Following
Refs. \onlinecite{zhang2009} and \onlinecite{wang2013}, a reasonable
estimate of the anisotropic $g$-factors would be $g_{x,\alpha} =
2.270$ and $g_{y,\alpha} = 2.269$. We set the hopping amplitude to
$t=0.8$ eV, while the transverse anisotropy parameter is set to $e =
5$ $\mu$eV, which is less than 1\% of the uniaxial anisotropy
parameter $d$. We do not consider coupling of the magnetic field to
the middle ion since it would change the energy levels of the high
energy sector by a very small amount, given that $b \ll U$. The goal
is to find $\phi$ and $b$ such that the in-plane anisotropy brought by
the external magnetic field compensates the intrinsic in-plane
anisotropy of the magnetic ions, removing the splitting between states
with maximum projection $|S_z|$, thus restoring degeneracy to the
ground state.

\begin{figure}[ht]
\includegraphics[width=0.9\columnwidth]{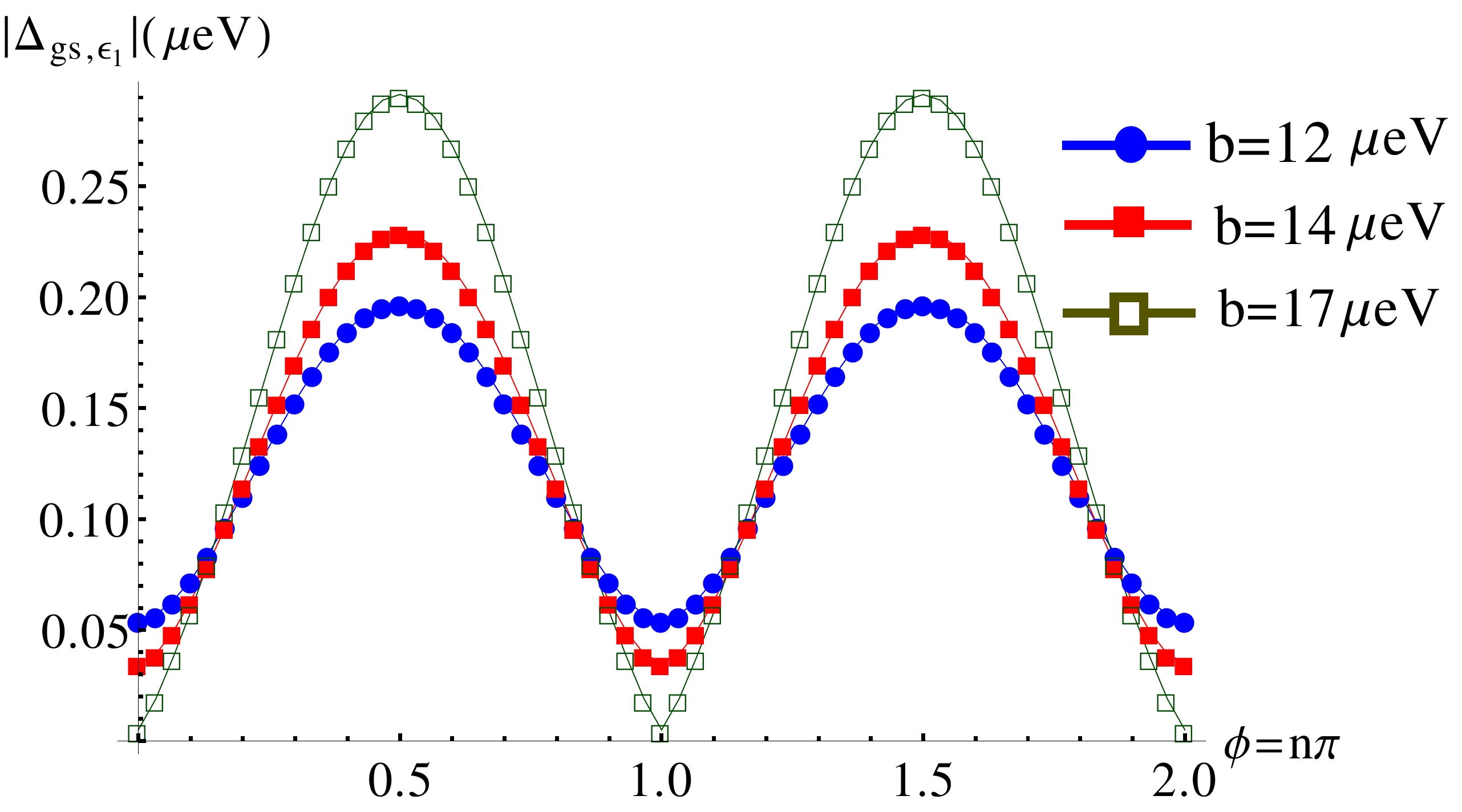}
\caption{(Color online) Splitting of the two lowest energy levels
  versus the transverse field angle $\phi$. The angle is measured with
  respect to the positive $x$ axis. We see that for various values of
  the magnetic field, the lowest splitting occurs at $\phi=\pi$.}
\label{fig:4}
\end{figure}

\begin{figure}[ht]
\includegraphics[width=0.93\columnwidth]{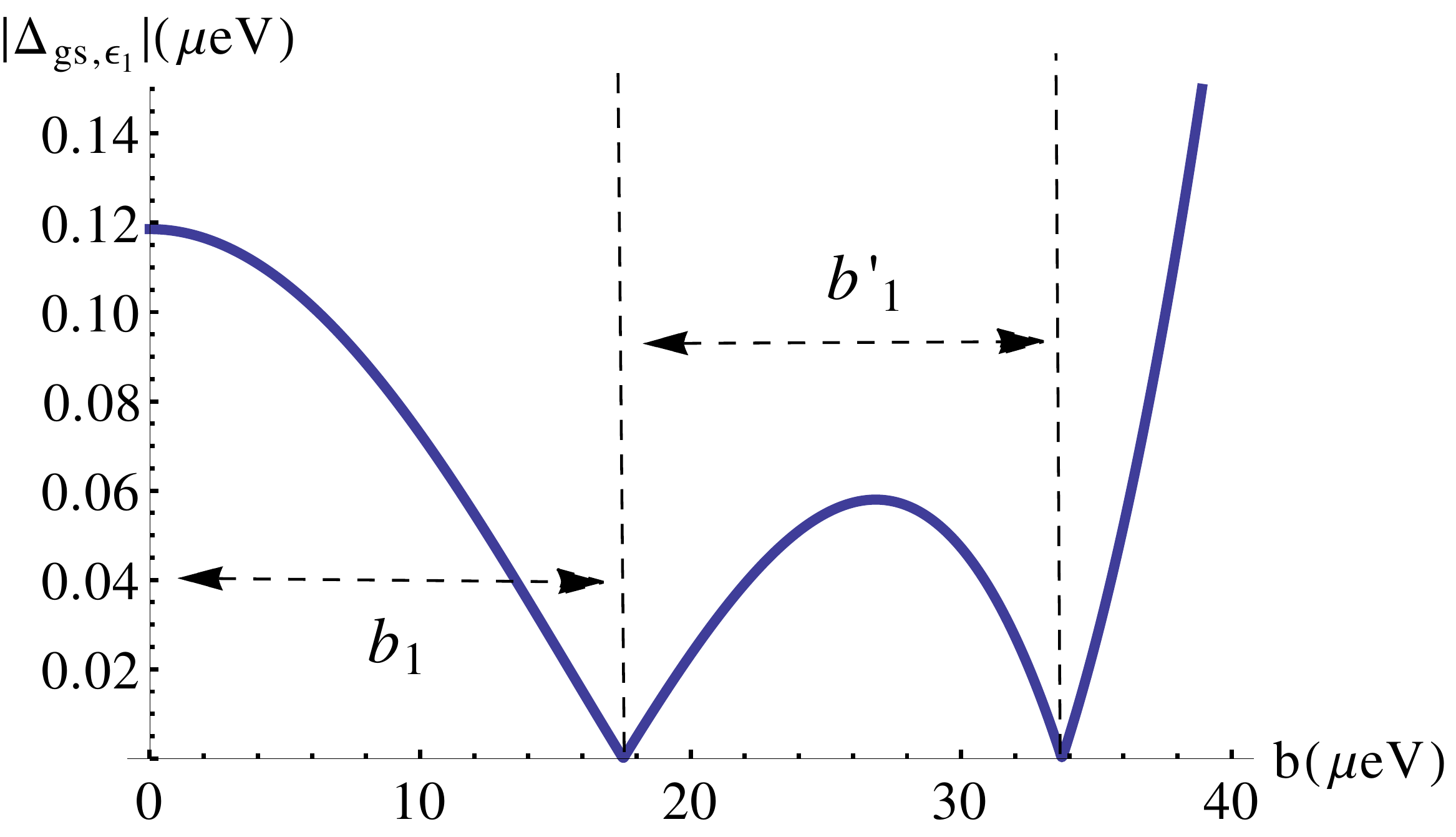}
\caption{(Color online) Splitting of the two lowest energy levels
  versus the magnitude of the magnetic field $b$ at an angle of $\phi
  = \pi$. We observe two degeneracy points where the splitting between
  the two lowest energy levels goes to zero. The first point occurs at
  $b\approx 18$ $\mu$eV ($B\approx 0.3$ tesla), while the second one
  occurs at $b\approx 34$ $\mu$eV ($B\approx 0.6$ tesla).}
\label{fig:5}
\end{figure}

In Fig. \ref{fig:4}, we show the resulting splitting of the two lowest
energy levels versus the angle $\phi$. The angle $\phi=\pi$ creates
symmetric spin paths for the tunneling of the molecule's
magnetization, leading to maximum destructive interference. In
Fig. \ref{fig:5} the splitting is shown as a function of the applied
field magnitude $b$ at this particular field direction
($\phi=\pi$). We observe that the second degeneracy point occurs
approximately at twice the value of the first one. We note that the
periodic modulation of the splitting with $b$ is as characteristic
feature of SMMs (see
 Refs. \onlinecite{friedman1996,wernsdorfer1999,vondelft+garg}).

\begin{figure}[ht]
\includegraphics[width=0.93\columnwidth]{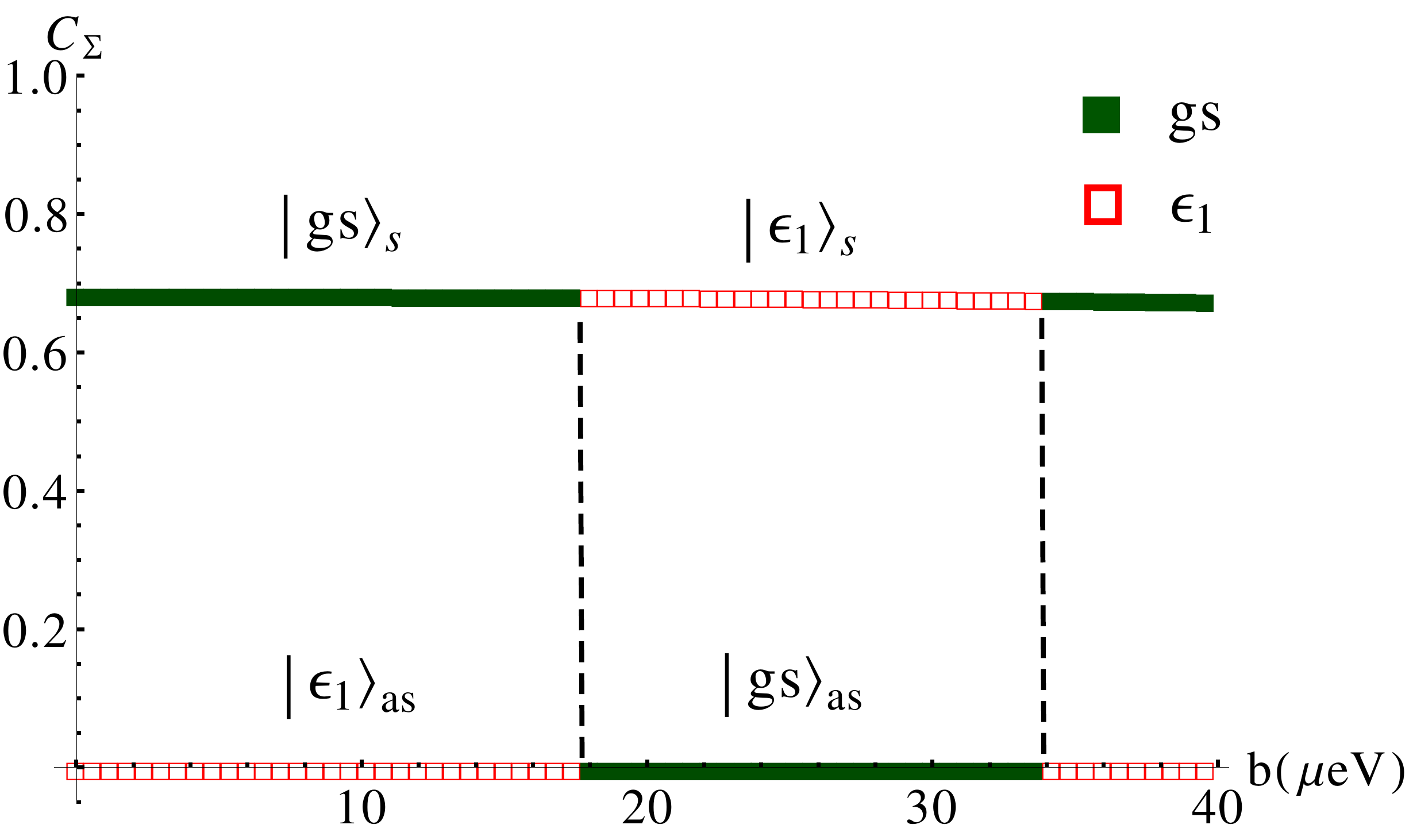}
\caption{(Color online) Symmetry of the ground state (filled squares)
  and first excited state (empty squares) versus the magnetic field
  magnitude $b$. At $b=0$ the ground and first excited states are
  $\vert \textrm{gs} \rangle = -0.68 (|+2 \rangle + |-2\rangle) +
  0.27\times (\mbox{contam.})$ and $\vert
  \textrm{\ensuremath{\epsilon}}_{1} \rangle = 0.69 (|+2 \rangle - |-2
  \rangle) + 0.22\times (\mbox{contam.})$, respectively, where
  $(\mbox{contam.})$ represents a contribution from spin states other
  than $|\pm 2\rangle$. As the transverse magnetic field grows, the
  amplitude $C_{\Sigma}$ tends to decrease, allowing for an increasing
  admixture with $S_{z}=0$ states. At the second degeneracy point
  (right dashed line), the symmetry of the states changes again.}
\label{fig:6}
\end{figure}

We now consider how the symmetry of the eigenstates is affected by
changes in the magnitude of the magnetic field. For this purpose, we
define the symmetry coefficient
\begin{equation}
\label{eq:28-1}
C_{\Sigma} =\frac{1}{2} \vert C_{1} + C_{2} \vert,
\end{equation}
where $C_{1,2}$ are the amplitudes of the two degenerate states along
the $S_z = \pm 2$ basis states. If $C_{\Sigma}=0$ the eigenstate is
antisymmetric, whereas if $C_{\Sigma}=C_{1,2}$ the eigenstate is
symmetric. The results are presented in Fig. \ref{fig:6}. We observe
that the ground state changes from symmetric to antisymmetric at the
first degeneracy point (left dashed line), while the first excited
state does the reverse. The energy splitting between the linear
combination of $S_{z}$ states (symmetric and antisymmetric) vanishes
at the degeneracy points.

\section{Sequential transport through the molecule}
\label{sec:transport}

While a fully-coherent transport approach can be used, it is
not essential in describing theoretically the effect of the transverse
magnetic field modulation on the SET I-V characteristics, which can be fully captured by a rate equation
approach in the dc limit. 

We study electronic transport in our model system by connecting the
molecule to two reservoirs of noninteracting electrons. The current
through the molecule can be controlled by applying a voltage
difference between the reservoirs, as well as by changing the total
charge of the molecule through an applied back-gate voltage. We define
a general charge-spin state by $(n,S)$ where $n$ and $S$ denote the
excess charge and the total spin of the molecule, respectively. For
simplicity, we only consider transitions between two charge states
($n=0,1$). Initially, the molecule is in the state $(0,2)$ where it
has a $3d^{8} - 2p^{6} -3d^{8}$ electronic configuration. We then
allow one electron to hop from the reservoirs into the molecule and
restrict it to be localized either on the $a_{1}$ or $c_{1}$ orbitals,
bringing the molecule to the $(1,\frac {3}{2})$ state, which comprises
both $3d^{9} - 2p^{6} - 3d^{8}$ and $3d^{8} - 2p^{6} - 3d^{9}$
electronic configurations. 

We assume that the coupling to the reservoirs is such that there is
equal probability for an electron to land in either one of the two
magnetic ions. Since one of the ions changes its oxidation state when
an electron is added, its anisotropy terms and the $g$-factors change
as well (see Refs.  \onlinecite{romeike2006a},
\onlinecite{romeike2006b}, \onlinecite{basler2005},
\onlinecite{feng2011}, and \onlinecite{zyazin2010}). 

Concerning model parameters, we assume a {\it reduction} in the local
uniaxial anisotropy of the ion upon changing its charge state and set
$d_{\alpha}^{\prime} = 0.3$ meV. It is worth noting that in the
context of the GSA, the anisotropy parameters change their values upon
varying the charge state of the molecule (see
Ref. \onlinecite{burzuri2012}). In our transport analysis,
$d_{\alpha}^{\prime}$ does not play an essential role since we will
focus on the contributions coming from the two lowest spin states of
the (1,3/2) charge-spin sector which can only have pure (or
combinations of) $S_{z}=\frac{3}{2}$ $z$-spin components. The modified
transverse anisotropy $e_{\alpha}^{\prime}$ does not take part in the
Hamiltonian for the $(1,\frac{3}{2})$ configuration since the total
spin of the ion receiving an electron becomes
$S_{\alpha}=\frac{1}{2}$, turning the local ground state of the ion a
Kramers doublet. Finally, for the sake of simplicity, we assume that
the same $g$ factors used for the $(0,2)$ configuration.

In order to evaluate single-electron transport, we start by
diagonalizing the Hamiltonian for different charge-spin
configurations. We then solve a set of coupled differential equations
for the time evolution of the quantum state probabilities of the
molecule. This approach is suitable for describing incoherent,
sequential transport across the molecule. In the context of our model,
its use is justifiable because the phase of the itinerant electron
does not play an essential role in the existence of degeneracy points
in the molecule's spectrum. In other words, the situation is similar
to that of a quantum dot where external parameters such as plunger
gate voltages can tune the states available to an itinerant electrons
tunneling in and out of the system regardless of the electron's phase
coherence. The only necessary assumption for the rate equation
approach to be valid is that the molecule should be weakly coupled to
the leads and the temperature sufficiently low. These conditions can
be cast as $\gamma, k_BT \ll |\Delta|$ (see notation below).

The rate equation describing the time evolution of the probability of
the molecule to be in an arbitrary state $m$ is given by
\cite{beenakker1991}
\begin{equation}
\label{eq:28}
\frac{dp_{m}}{dt} = -p_{m} \sum_{m^{\prime}} \Gamma_{m\rightarrow
  m^{\prime}} + \sum_{m^{\prime}} p_{m^{\prime}}\,
\Gamma_{m^{\prime}\rightarrow m},
\end{equation}
where $\Gamma_{m \rightarrow m^{\prime}}$ is the transition rate
between different eigenstates $m$ and $m^{\prime}$. The first term on
the r.h.s of Eq.(\ref{eq:28}) comprises outgoing terms describing the
transition from an initial state $m$ to a final state $m^{\prime}$ by
either taking away or adding an extra electron. The second term
represents processes where a final state $m^{\prime}$ transitions back
to an initial state $m$ (again, by means of either taking away or
adding an extra electron). Equation (\ref{eq:28}) can be written in
matrix form as
\begin{equation}
\label{eq:17-1}
\frac{d{\bf P}}{dt} = {\bf \Gamma} \cdot {\bf P},
\end{equation}
where ${\bf P}$ is a vector containing the probabilities $\{p_{m}\}$
and ${\bf \Gamma}$ is a matrix with entries $\{\Gamma_{m
  m^{\prime}}\}$ which are related to the the transition rates in the
following way:
\begin{eqnarray}
\label{eq:18-1}
\Gamma_{mm^{\prime}} & = & \Gamma_{m^{\prime}\rightarrow m} \qquad
\textrm{if} \qquad m \ne m^{\prime} \nonumber \\ \Gamma_{mm} & = & -
\sum_{m\ne m^{\prime}} \Gamma_{m \rightarrow m^{\prime}}.
\end{eqnarray}

In the stationary regime, we look for the steady-state probabilities
$p_{m}$ such that $dp_{m}/dt = 0$. In matrix form, this is achieved by
finding the eigenvector ${\bf P}$ of the matrix ${\bf \Gamma}$ with a
zero eigenvalue. The transition rates from a state $m$ to a state
$m^{\prime}$ when adding (or subtracting) electrons ($e^{-}$) to
(from) the molecule are given by
\begin{equation}
\label{eq:31}
\Gamma_{m \rightarrow m^{\prime}}^{\tau} = \left\{ \begin{array}{ll}
  \gamma^{\tau}_{m^{\prime},m}\, f_{\tau} (\mu_{m^{\prime} m} -
  eV_{\tau}), & e^{-}\ \textrm{in},
  \\ \gamma^{\tau}_{m,m^{\prime}}\ \left[ 1 - f_{\tau} (\mu_{m
      m^{\prime}} - eV_{\tau}) \right], &
  e^{-}\ \textrm{out} \end{array} \right.
\end{equation}
where the index $\tau=L,R$ refers to the left and right reservoirs,
respectively. Here, $f_{\tau}(x) = 1/(1+e^{x/k_BT})$ is the Fermi
distribution of the left ($L$) or right ($R$) reservoirs. The
electrochemical potential of the molecule, i.e., the energy required
to go from a state $m$ to state a $m^{\prime}$, is defined as
\begin{equation}
\label{eq:33-1}
\mu_{m m^{\prime}} = \epsilon_{m} - \epsilon_{m^{\prime}} - e\eta
V_{g},
\end{equation}
where $\epsilon_{m}$ and $\epsilon_{m^{\prime}}$ are the energy
eigenvalues corresponding to eigenstates $m$ and $m^{\prime}$,
respectively, and $V_g$ is the backgate voltage (we set the lever arm
coefficient $\eta=1$). The coefficients
$\gamma^{\tau}_{\beta,\beta^{\prime}}$ are the tunneling rates between
two eigenstates $\beta$ and $\beta^{\prime}$ of the molecule and are
given in the Golden rule approximation by
\begin{equation}
\label{eq:34}
\gamma^{\tau}_{\beta, \beta^{\prime}} = 2\pi \rho \sum_{\sigma} \vert
T_{\beta \beta^{\prime}}^{\sigma,\tau} \vert^{2}
\end{equation}
where $\rho$ is the density of states of both reservoirs (here
considered constant) and
\begin{equation}
\label{eq:36}
T_{\beta \beta^{\prime}}^{\sigma,\tau} = \sum_{j}t_{j,\tau} \langle
\beta \vert c_{j\sigma}^{\dagger} \vert \beta^{\prime} \rangle
\end{equation}
denote the tunneling matrix elements. The operator
$c_{\alpha_{j}\sigma}^{\dagger}$ creates an electron with spin
$\sigma$ on the single-particle orbital $j$ of ion $\alpha$. To
simplify the calculations we assume that $t_{j,\tau}$ is independent
of the orbital, thus we drop the $j$ index, also we assume that
$t_{L}=t_{R}$. Since different charge configurations of the molecule
involve states with different spin quantum numbers, selection rules
for spin transitions become important in the transport
calculations. The tunneling matrix elements satisfy the selection
rules $\vert S_{m} - S_{m^{\prime}} \vert = 1/2$, $\vert S_{zm} -
S_{zm^{\prime}} \vert = 1/2$ and $\vert N_{m} - N_{m^{\prime}} \vert =
1$, for any two eigenstates of the Hamiltonian belonging to different
charge-spin sectors. The net steady-state current through the left
lead (we omit the $L$ subscript) is given by
\begin{equation}
\label{eq:37}
I = \vert e \vert \sum_{m m^{\prime}} \Gamma_{m \rightarrow
  m^{\prime}}^{L} p_{m},
\end{equation}
where $m,m^{\prime}$ are indexes for different charge states. In
Eq. (\ref{eq:37}), multiply the r.h.s it by $-1$ if an electron is
going out of the molecule and into the left reservoir.

\section{Effect of spin tunneling modulation on transport}
\label{sec:modulation}

\begin{figure}[ht]
\includegraphics[width=0.9\columnwidth]{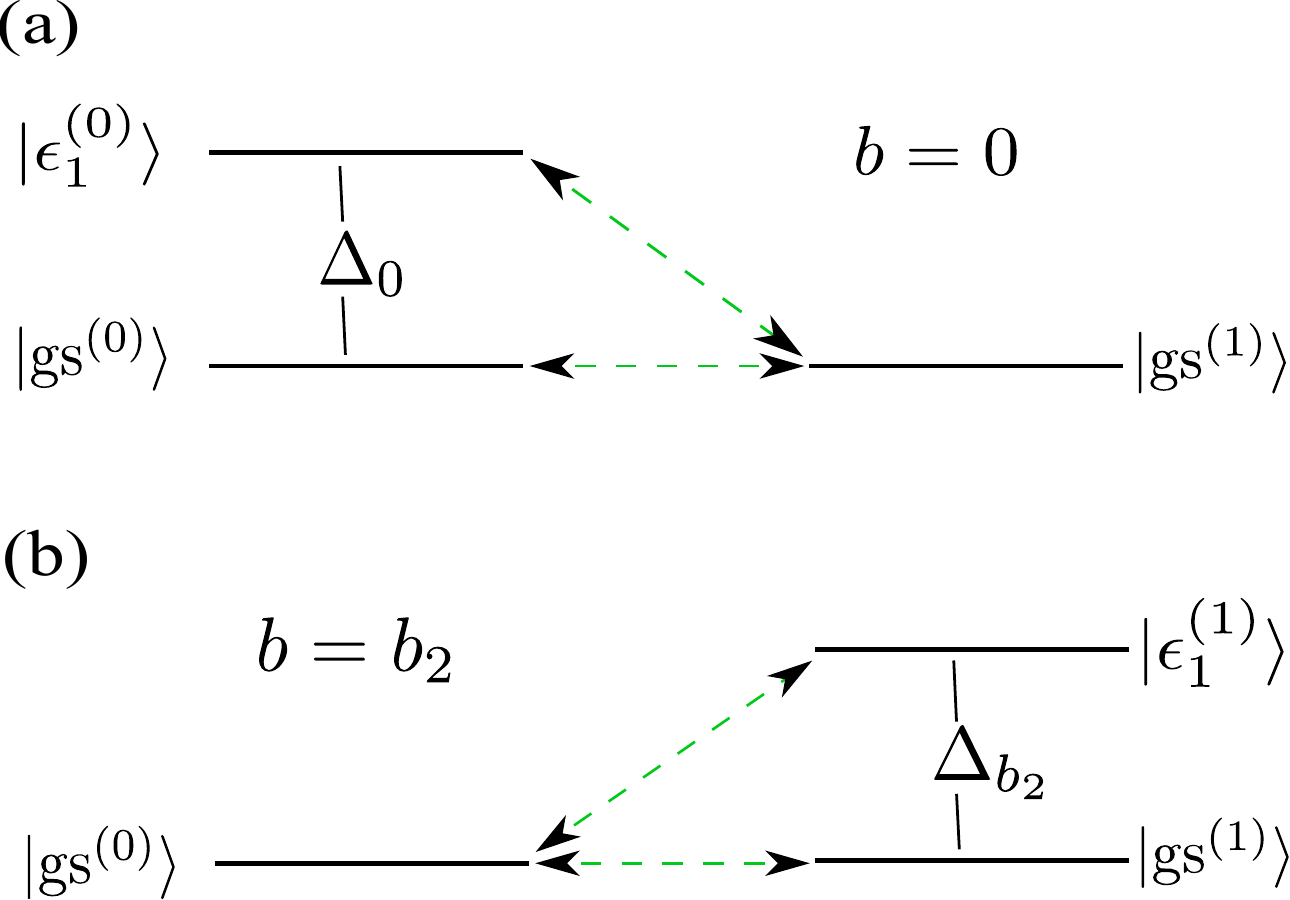}
\caption{(Color online) Allowed transitions (dashed arrows) between
  the two lowest energy levels for the $(0,2)$ and $(1,3/2)$
  charge-spin sectors. In (a), $\Delta_0$ is the splitting between
  eigenstates for zero field ($b=0$) while in (b) $\Delta_{b_2}$ is
  the splitting when the field is tuned to the degeneracy point
  ($b=b_{2}$).}
\label{fig:7}
\end{figure}

Transport is evaluated for different values of the field between in
the range $[0:b_2]$, where $b_{S}$ denotes the magnitude of the field
for which a degeneracy point occurs for the charge-spin sector with
total spin $S=2$. In order to see signatures of magnetization
tunneling interference on transport, we only consider transitions
between the two lowest energy levels of each charge-spin sector, as
shown in Fig. \ref{fig:7}. The ground and first excited eigenstates of
charge sector $n$ are labeled as $\vert \textrm{gs}^{(n)}\rangle$ and
$\vert \epsilon_{1}^{(n)}\rangle$, respectively. In Fig. \ref{fig:7},
we see that for $b=0$ the eigenstates of the $(0,2)$ configuration are
linear combinations of $\vert S_{z} =\pm 2 \rangle$ spin states, while
the ground state of the $(1,\frac{3}{2})$ sector is two-fold
degenerate with the $\vert S_{z} = \pm \frac{3}{2} \rangle$ spin
states having the same energy. For $b=b_{2}$ the tunnel splitting of
the $(0,2)$ configuration goes to zero and the ground state becomes
the two-fold degenerate $\vert S_{z} = \pm 2 \rangle$ spin states. As
for the $(1,\frac{3}{2})$ sector, an energy splitting
($\Delta_{b_{2}}$) is induced by the magnetic field, with the
eigenstates being now a linear combination of $\vert S_{z} =\pm
\frac{3}{2} \rangle$ spin states. In both cases, the corresponding
selection rules allow transitions between all eigenstates.

Interesting physics can also be found if all combinations of $S_{z}$
states corresponding to a particular total spin $S$ of the molecule
are considered. This is because anisotropy contributions tend to
contaminate the lowest energy levels with a small admixture of
$S_{z}=0$ states, opening transitions in the transport that would
otherwise have been forbidden by selection rules. A study of these
contributions to transport within the GSA can be found in
Ref. \onlinecite{romeike2006b}. Since we are primarily interested in
how the modulation of the transverse anisotropy splitting affects
transport, we only consider the two lowest energy $S_{z}$ states. For
other states, the energy cost to access different transitions would be
too large compared to the energy gap generated by the transverse
anisotropy.

\begin{figure}[t]
\includegraphics[width=1\columnwidth]{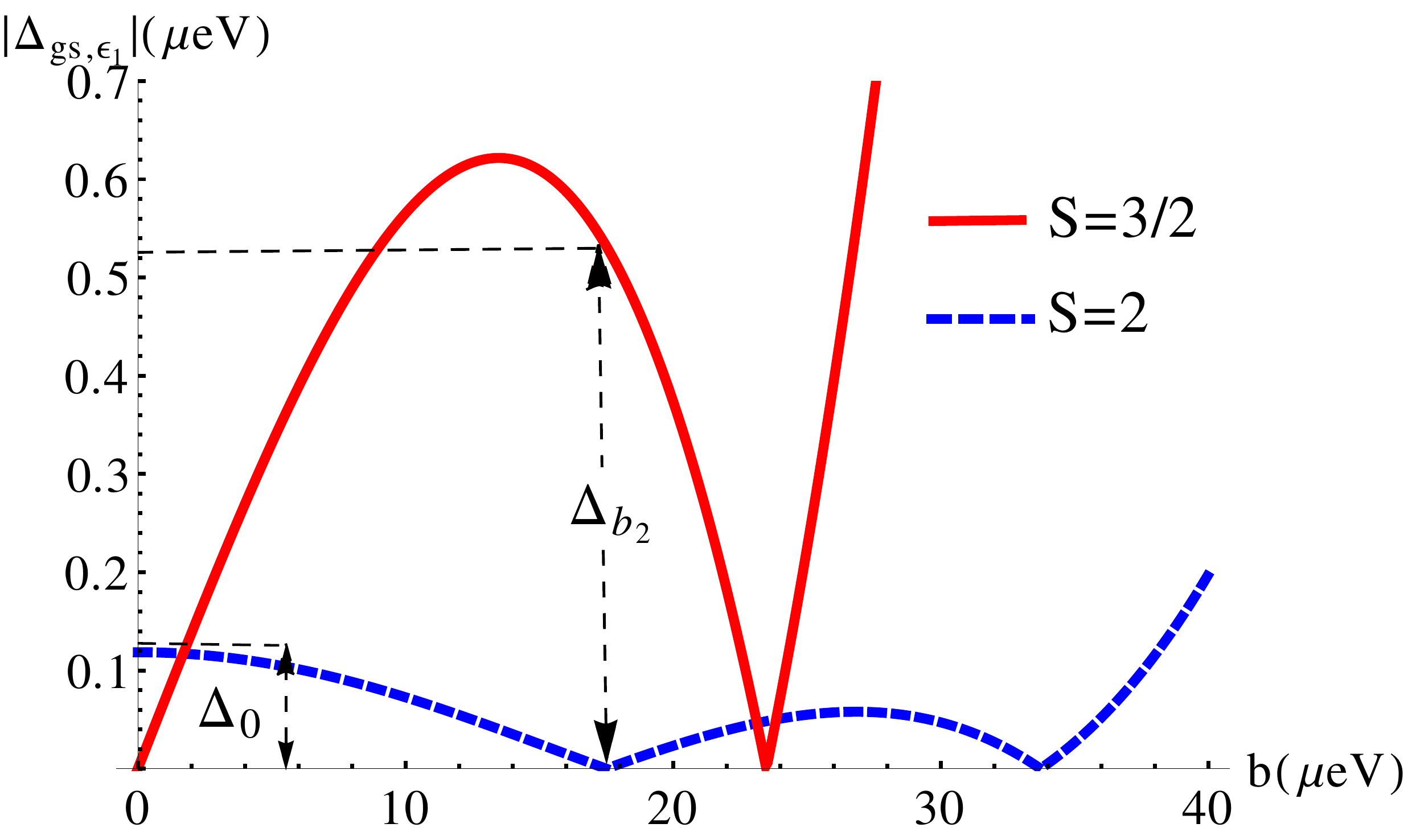}
\caption{Splittings between the two lowest energy levels for the
  $S=3/2$ (solid line) and $S=2$ (dashed line) total spin sectors. The
  first degeneracy point for $S=2$ is at $b_{2} \approx 18\ \mu$eV
  (equivalent to 0.3 tesla).}
\label{fig:8}
\end{figure}

We tune the gate voltage $V_{g} $ so that the ground states energies
for the $(0,2)$ and $(1,\frac{3}{2})$ charge-spin sectors are aligned
and vary $V_g$ slightly around this point. Since the tunnel splitting
due to transverse anisotropies is very small, of the order of the
$\mu$eV (see Fig. \ref{fig:8}), a very small temperature as well as a
very low coupling $\gamma_{\beta,\beta^\prime}^\tau$ are required to
resolve features in the electronic transport that can be associated to
the tunneling of the magnetization. Thus, we set the temperature in
the reservoirs to $T=0.1$ mK and choose the product
$\rho|T_{\beta\beta^\prime}^\tau|^2$ so that both $k_BT$ and
$\gamma_{\beta\beta^\prime}^\tau$ are much smaller than the energy
level separation within the molecule. We note that while, in practice,
a small value for $\gamma_{\beta,\beta^\prime}^\tau$ can be achieved
by chemically engineering the SMM ligands, arriving at such low
temperatures in single-electron transistor setups is quite
challenging.

\begin{figure}[ht]
\includegraphics[width=0.92\columnwidth]{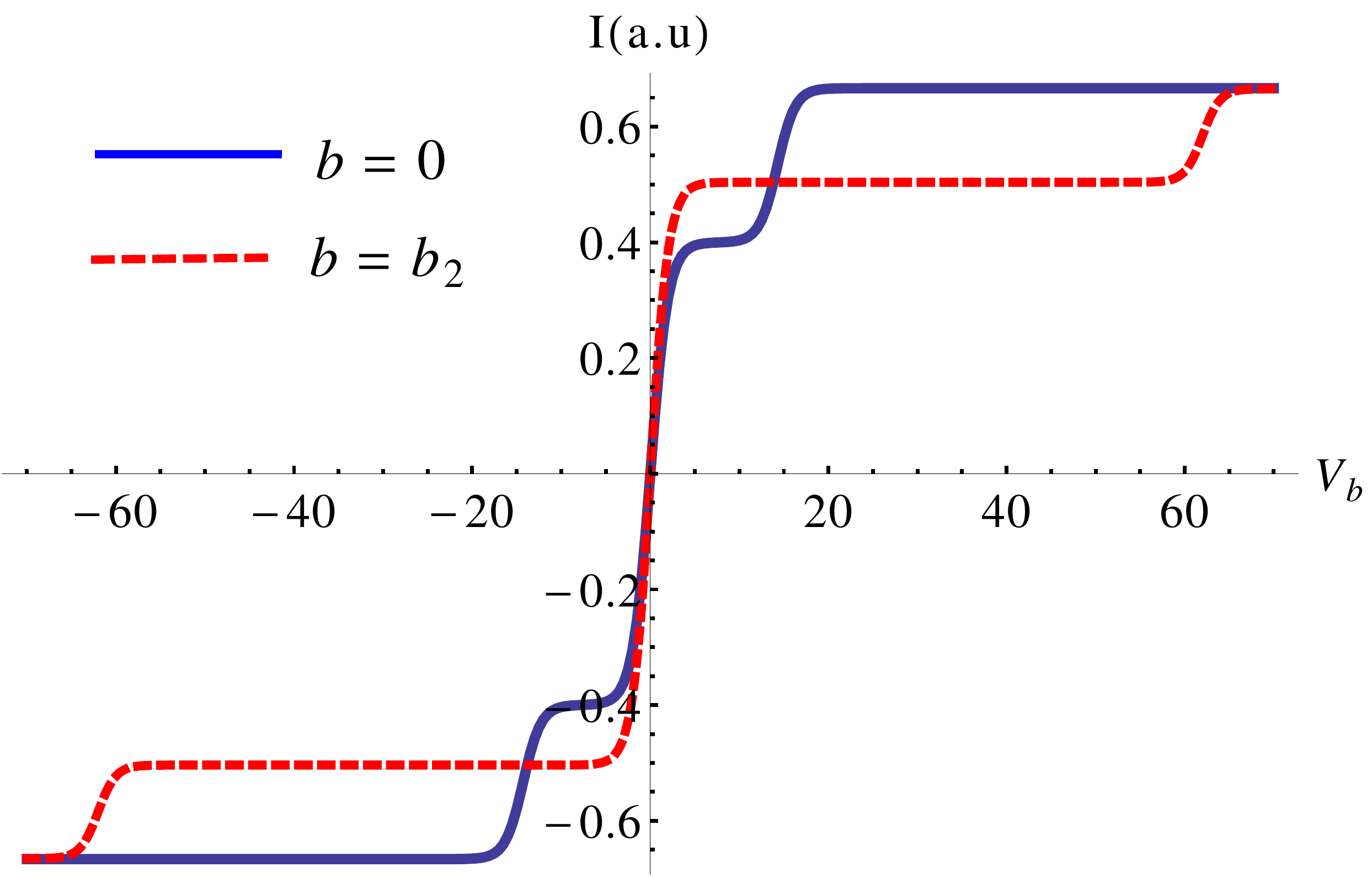}
\caption{Current through left lead versus bias voltage at $b=0$ (solid
  line) and $b=b_{2}$ (dashed line) for $T=1$ mK (current is shown in
  arbitrary units). The bias voltage is plotted in units of
  $k_{B}T/\vert e \vert$. The gate voltage for each curve is such that
  the ground states for the two charge sectors involved are aligned.}
\label{fig:9}
\end{figure}

\begin{figure}[hb]
\includegraphics[width=0.95\columnwidth]{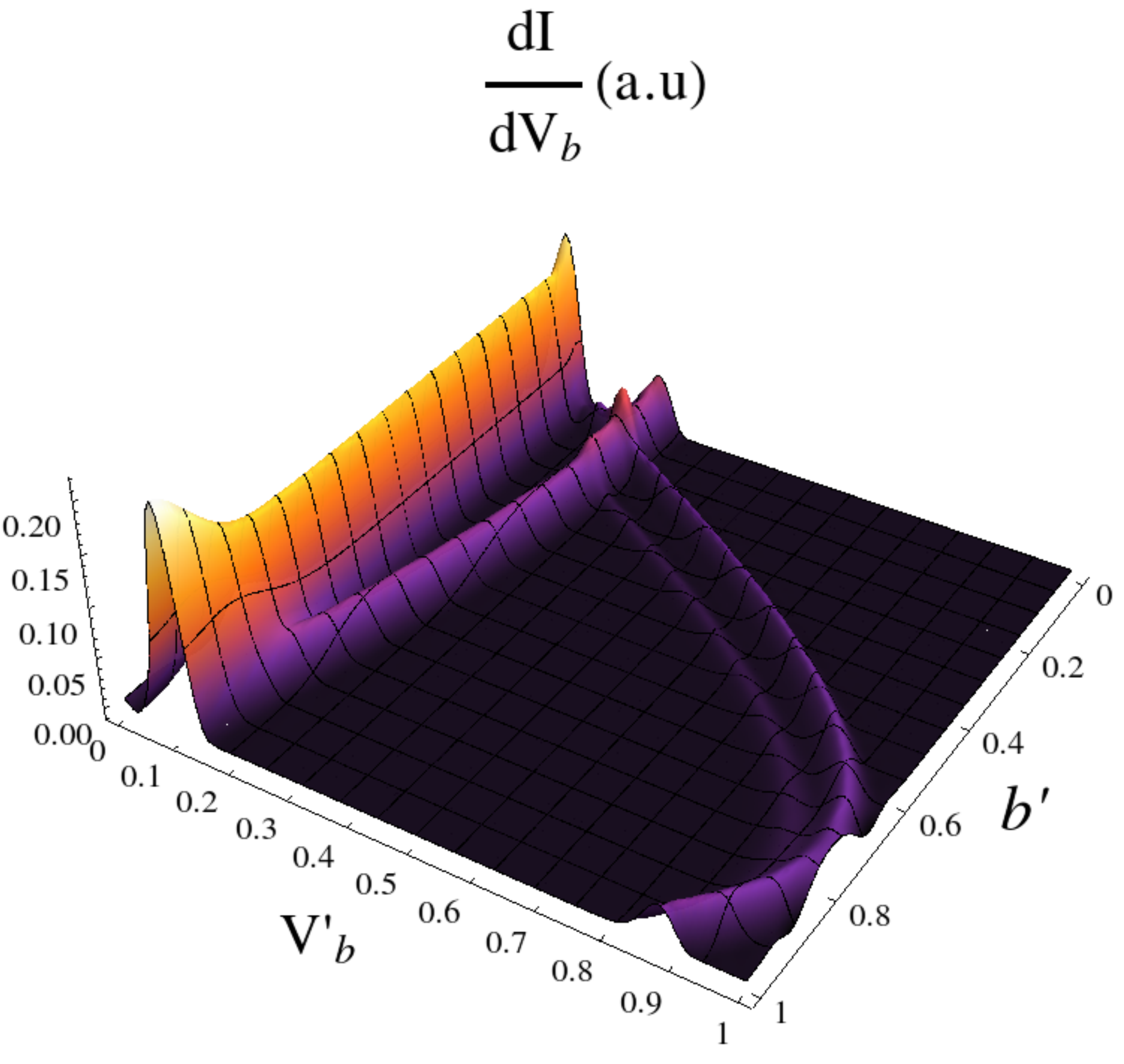}
\caption{Plot of the differential conductance $dI/dV_{b}$ as a
  function of the dimensionless bias voltage $V_b^\prime =
  V_{b}/V_{\textrm{max}}$ (here $V_{\textrm{max}}=75 k_{B}T/\vert e
  \vert$) and the dimensionless magnetic field $b^\prime =
  b/b_{2}$. The gate voltage is varied so that the ground states of
  the two charge sectors, $(0,2)$ and $(1,\frac{3}{2})$, are kept
  aligned. The differential conductance is given in arbitrary units.}
\label{fig:10}
\end{figure}

\begin{figure}[ht]
\includegraphics[width=0.91\columnwidth]{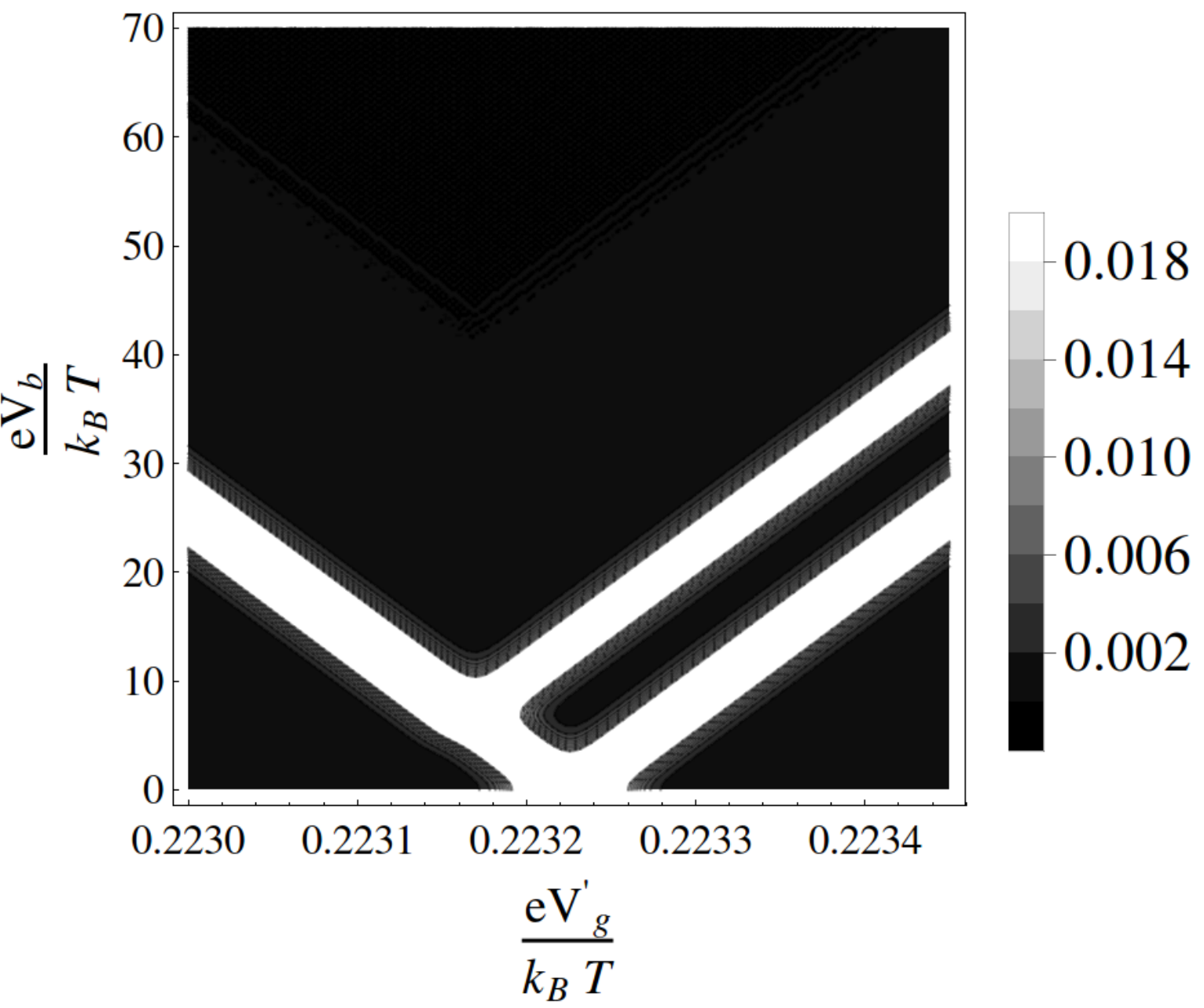}
\caption{Contour plot of the differential conductance $dI/dV_{b}$ as a
  function of bias and gate voltage at $b=0$.}
\label{fig:11}
\end{figure}

\begin{figure}[t]
\includegraphics[width=0.91\columnwidth]{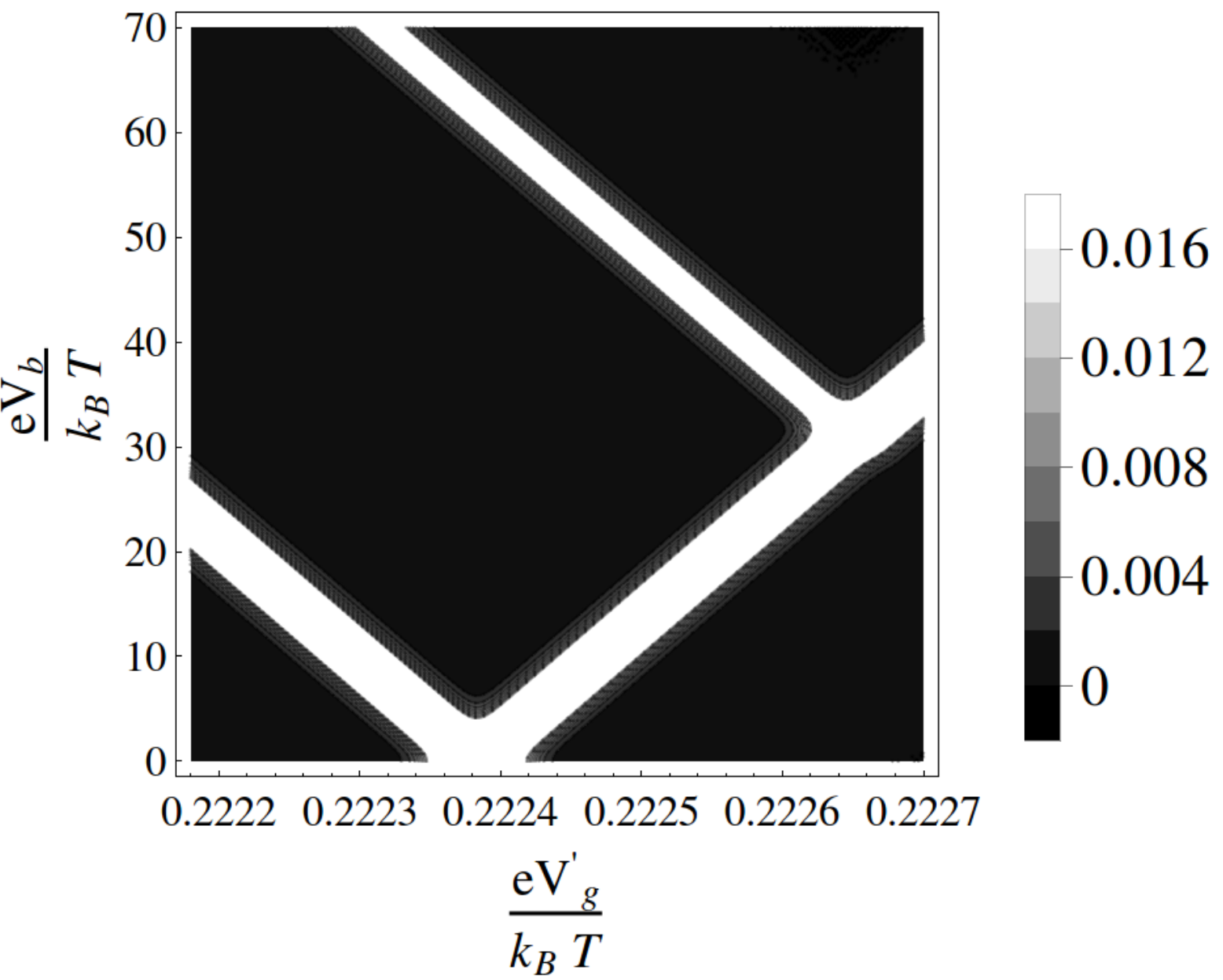}
\caption{Contour plot of the differential conductance, $dI/dV$ in
  arbitrary units, as a function of bias and gate voltage at
  $b=b_{2}$.}
\label{fig:12}
\end{figure}

The left-lead current ($I$) to/from the molecule is shown in
Fig. \ref{fig:9}. For $b=0$, transitions between excited and ground
states are seen in the current steps at zero bias voltage and at
$V_{b}/k_BT \approx \pm 13$. When $b=b_{2}$, these steps can be seen
instead at $V_{b}/k_BT \approx \pm 61$. Figure \ref{fig:10} shows a
plot of the differential conductance ($dI/dV_{b}$) as a function of
the bias voltage and the magnetic field. Resonance peaks seen at
$b^{\prime}=0$ and $b^{\prime}=1$ correspond to the current steps of
Fig. \ref{fig:9}. As one approaches the degeneracy point
($b^{\prime}=1$), the conductance peak corresponding to the $\vert
\textrm{gs}^{(1)} \rangle \leftrightarrow \vert \epsilon_{1
}^{(0)}\rangle$ transition is shifted towards the larger conductance
peak (at $V^{\prime}_{b}=0$). At the degeneracy point, this peak is
absorbed by the zero-bias conductance peak, increasing the current
flow between the ground states of the molecule. In addition, we notice
that the conductance decreases when the field is close to zero. We
also observe that new resonances appear as a consequence of the
field-induced energy gap of the $(1,\frac{3}{2})$ charge-spin
sector. These can be seen in the conductance peaks coming out of the
zero-bias and $\vert \textrm{gs}^{(1)} \rangle \leftrightarrow \vert
\epsilon_{1 }^{(0)}\rangle$ transition resonance peaks in the
$b^{\prime}=0$ plane. For some values of the field, conductance peaks
arise as resonances merge with each other. These peaks appear when the
electrochemical potentials in the dot are the same i.e.,
$\mu_{\textrm{gs}^{(0)},\epsilon^{(1)}_{1}}=\mu_{\epsilon^{(0)}_{1},\epsilon^{(1)}_{1}}$,
$\mu_{\textrm{gs}^{(0)},\textrm{gs}^{(1)}}=\mu_{\epsilon^{(0)}_{1},\epsilon^{(1)}_{1}}$,
and
$\mu_{\textrm{gs}^{(1)},\epsilon^{(0)}_{1}}=\mu_{\epsilon^{(0)}_{1},\epsilon^{(1)}_{1}}$. Figures
\ref{fig:11} and \ref{fig:12} show contour plots of the differential
conductance versus bias and gate voltage. The positive slope line seen
for $b=0$ in Fig. \ref{fig:11} corresponds to the $\vert
\textrm{gs}^{(1)} \rangle \leftrightarrow \vert \epsilon_{1
}^{(0)}\rangle$ transition and is eliminated upon tuning the field to
the degeneracy point. At the same time, the applied magnetic field
creates an energy gap allowing the $\vert \textrm{gs}^{(0)} \rangle
\leftrightarrow \vert \epsilon_{1 }^{(1)}\rangle$ transition, which
corresponds to the negative slope line in Fig. \ref{fig:12}.

In Figs. \ref{fig:11} and \ref{fig:12}, in order to show in detail the
fine features that appear upon crossing a degeneracy point, the gate
voltage $V_g$ was shifted and rescaled: $V_{g}^{\prime}= (V_{g}-V_{s})
\times 10^{3}$, where $eV_{\textrm{s}} = 8.169$ eV.

\section{Conclusion}
\label{sec:conclusion}

We have studied the incoherent electronic transport through an
anisotropic magnetic molecule using a microscopic model that provides
spectral properties similar to those of a multi-ion single-molecule
magnet. By describing the molecule's core as a set of a few
multi-orbital quantum dots, we open the door to a better understanding
of the interplay between internal degrees of freedom of the molecule
and its transport properties. This cannot be done within the giant
spin approximation which is usually the starting point for
characterizing the behavior of SMMs.

Another advantage of our model is its simplicity: by reducing the
degrees of freedom to a manageable number, the model makes it possible
to study in more detail the influence of the molecule's geometry and
ion arrangement, as well as of the electron path across the molecule,
on transport. In addition, by keeping the number of degrees of freedom
small, it might be possible to go beyond the incoherent regime and use
this model to study strongly correlated phenomena such as the Kondo
effect. These are currently out of the reach of fully atomistic
calculations, such as those of Refs. \onlinecite{park2010} and
\onlinecite{nossa2012}.

Our model captures the essential phenomenology of a SMM, including the
quantum tunneling of the net magnetization. We have illustrated this
point by showing a modulation of the non-linear I-V characteristics
upon the application of an transverse magnetic field. The appearance
and disappearance of resonance lines is a clear indication of the
existence of degeneracy points in the molecule's spectrum at certain
values and directions of the transverse field. This behavior is
similar to what is expected for a SMM in the giant spin approximation,
where the destructive interference between tunneling trajectories of
the giant spin create a periodic dependence on the transverse
field. However, the lack of a well-defined topological phase in our
model prevents us from making a direct connection between the
modulation we observe and Berry phase interference.

Our model does have some limitations. For instance, we restrict the
orbitals that participate in the hopping terms of the molecule's
Hamiltonian. In addition, not all configurations are included and the
interactions with ligands is only taken into account in at a
phenomenological level.

The energy spectrum in our model is very sensitive to the choice of
parameters and there is a complex interplay between the different
interactions present in the model. In general, on-site anisotropies,
Coulomb and overlap interactions intra and inter transition metals are
not trivial to estimate. We have tried to use realistic or reasonable
values whenever possible. These parameters depend on the electronic
structure of the magnetic ions as well as on the geometrical
configuration of the ligands surrounding the magnetic core. The bond
angle between two magnetic ions is also of key importance in our
calculations. For the particular three-ion model we studied, we chose
a right bond angle to be 90 degrees and magnetic ions with parallel
anisotropies. This allowed us to neglect the sigma overlap between the
$d_{x^2-y^2}$ orbitals of the magnetic ions and the $p_{z}$ orbital of
the bridging diamagnetic ion, as well as to prevent any contamination
by $d_{3z^2-r^2}$ orbitals. The result was an effective superexchange
ferromagnetic interaction that competed with the $S_{z}=0$ ground
state favored by the local uniaxial anisotropies. Even with all these
constraints, we found that the model Hamiltonian yielded a high-spin,
tunneling-split ground state, as it is typical for SMMs. A crucial
parameter for our study is the on-site in-plane anisotropy $e$, which
controls the splitting and competes with the in-plane magnetic field.

Our calculations show that at very low temperatures certain
transitions are suppressed when a transverse magnetic field is tuned to
a special direction and value that make the ground state twofold
degenerate. This has a direct effect on non-equilibrium electronic
transport across the molecule. At small enough bias voltages, the
effect of the external field when tuned in-between the degeneracy
points is to modulate the access to excited states. This in turn
shifts the peaks in the $dI/dV$ response in an amount proportionally
to the tunnel splitting of spin states.

The fact that we have to rely on very low temperature (in the
mili-Kelvin range) to visualize these features indicates that they
will be challenging to observe experimentally. Molecules with a large
tunnel splitting are better suited for exploring the interplay between
transport and quantum tunneling of the magnetization. This will
require a relatively strong in-plane anisotropy as compared to the
uniaxial anisotropy, which goes somewhat against the usual synthesis
effort, which aims at increasing the uniaxial anisotropy. It is worth
noting that recent advances in experimental setups indicate that small
splittings may be observable.\cite{urdampilleta2011,burzuri2012}

We plan to further explore our microscopic formulation to consider
more complex molecules with multi-ion transition-metal cores. We are
currently extending our approach to investigate field modulation
effects in the electronic transport of mononuclear SMMs, which consist
of one rare-earth ion within a ligand cage. These systems have been
synthesized with success in recent years (see for instance
Refs. \onlinecite{aldamen2008,aldamen2009,titel2011,martinezperez2012}).

\begin{acknowledgments}

This work was supported by NSF ECCS under Grant No. 1001755. The
authors would like to thank Enrique del Barco, George Martins, and
Edson Vernek for valuable comments.

\end{acknowledgments}



\end{document}